\documentclass[aps,preprint,amssymb,12pt,floatfix]{revtex4}
\usepackage{subfigure}
\usepackage{graphicx}
\usepackage{rotating}
\usepackage{tabularx}
\usepackage{wrapfig}

\def\la{\langle}
\def\ra{\rangle}

\begin{document}

\renewcommand{\thefootnote}{\fnsymbol{footnote}}

\title{Influence of Nanoparticle Size and Shape on Oligomer Formation of an Amyloidogenic Peptide}

\author{Edward P. O'Brien$^{1,2}$\footnote{Current address: Dept. of Chemistry, University of Cambridge, Cambridge CB2 1EW, UK.}, John. E. Straub$^{3}$, Bernard R. Brooks$^{2}$ and D. Thirumalai$^{1,4}$\footnote{Corresponding author:
Institute for Physical Science and Technology, University of Maryland,
College Park, MD 20742, phone: 301-405-4803; fax: 301-314-9404;
e-mail: thirum@umd.edu}}
\affiliation{\small $^1$Biophysics Program, Institute for Physical Science and Technology,\\
University of Maryland, College Park, MD 20742\\
$^2$Laboratory of Computational Biology\\
National Heart Lung and Blood Institute\\
National Institutes of Health, Bethesda, MD 20892\\
$^3$Department of Chemistry,\\
Boston University, Boston, MA 02215\\
$^4$Department of Chemistry and Biochemistry,\\
University of Maryland, College Park, MD 20742}
\vspace{2cm}
\date{\today}

\maketitle

\noindent
\textbf{Abstract}
Understanding the influence of macromolecular crowding and nanoparticles on 
the formation of in-register $\beta$-sheets, the primary structural component
of amyloid fibrils, is a first step towards describing  
\emph{in vivo} protein aggregation 
and interactions between synthetic materials and proteins. 
Using all atom molecular simulations in implicit solvent we illustrate the effects of nanoparticle 
size, shape, and volume fraction on oligomer formation 
of an amyloidogenic peptide from the transthyretin protein.
Surprisingly, we find that 
inert spherical crowding particles destabilize in-register $\beta$-sheets formed
by dimers while stabilizing $\beta$-sheets comprised of trimers and tetramers. 
As the radius of the nanoparticle increases  
crowding effects decrease, implying smaller crowding particles have the largest influence
on the earliest amyloid species.
We explain these results using a theory based on the depletion effect.
Finally, we show that spherocylindrical crowders destabilize the ordered $\beta$-sheet dimer to a greater extent than
spherical crowders, which underscores the influence of nanoparticle shape on protein aggregation. 

\begin{center}
\includegraphics[width=5.1cm]{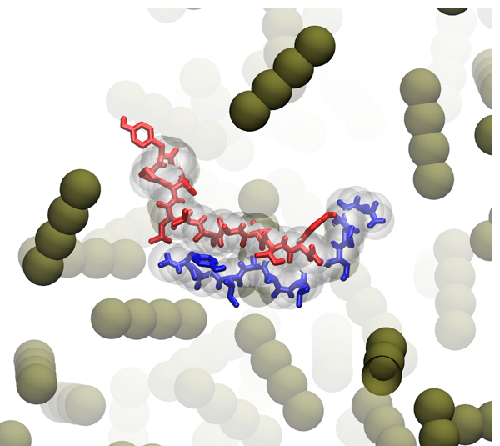}
\\TOC Graphic
\end{center}

\begin{center}
\end{center}

\newpage

\renewcommand{\figurename}{FIG.}

Molecular crowding can
have a profound effect on virtually all biological processes such as protein folding, viral capsid assembly and
protein aggregation \cite{EatonPNAS1976,Ross77JMB,MintonARB2008,thirumPNAS2005,Mateu2005,Straub10COSB,StaggPNAS2007,ZhouCR2009,CheungPNAS2010,Shea11COSB}.
It is estimated that 20 to 30\% of a typical cell's volume 
is occupied by DNA, protein, lipids and sugars \cite{MintonARBPBMS93,MintonARB2008,ElcockCOSB2010}. Such a crowded environment restricts the conformations explored by biological macromolecules, thus affecting the balance of thermodynamic forces that help
regulate cellular processes.
More recently, it has also been realized that synthetic nanoparticles (NPs) such as quantum dots, carbon nanotubes, and gold and 
other colloidal particles also affect the stability and function of proteins 
\cite{Linse07PNAS,Auer09PlosCompBiol,Colvin07PNAS}. Understanding nanoparticle-protein 
interactions are important in our ability to use NPs for drug delivery and controlling environmental toxicity \cite{Colvin03NatureBiotech}.   

In the context of  aggregation of amyloidogenic peptides, which is the focus of the present study, NPs  can greatly influence the stabilities of the molecular species that accumulate along the routes to fibril formation.   Amyloid fibrils are experimentally
characterized as cross-$\beta$ structures that are
rich in $\beta$-sheet content \cite{ChanBIOCHEM2005,DobsonARB2006,EisenbergCOSB2006}.
Formation of amyloid is linked to  a number of  human diseases \cite{KellyNature2002,DobsonARB2006},
and it appears that almost any polypeptide chain  can form amyloid under appropriate
solution conditions \cite{DobsonARB2006}. Even small peptides \cite{EisenbergNature2005}, such as the  
fragment comprised of residues 105-115 from the Transthyretin (TTR) protein, 
form amyloid fibrils \cite{GriffinPNAS2002,GriffinPNAS2004}. TTR fibrils are
associated with senile systemic amyloidosis and familial amyloid polyneuropathy I \cite{KellyNature2002}.  

General theoretical arguments based on the depletion effect \cite{OosawaJCP1954,ShawPRA1991,Fleer08AdvCollIntSci,OdijkJSB2001,MintonARB2008,thirumPNAS2005} suggest 
that monodisperse spherical NPs should promote protein aggregation. The depletion effect, which results in non-specific entropy-induced attraction between proteins in the
presence of non-adsorbing NPs, 
arises from the volume excluded to the NPs by the monomer subunits of aggregating proteins. 
Consider for example two proteins that are spherical with a radius $R_g^N$. When two such 
proteins approach each other up to a distance on the order of the radius, $R_C$, of the crowding particles, 
the particles can no longer fit and are therefore expelled from the gap formed by the proteins. Thus, the crowding particles exert
an osmotic pressure on the associating monomers resulting in a net inter-protein attraction \cite{OosawaJCP1954,ShawPRA1991,Fleer08AdvCollIntSci}.
As a result, inert crowding particles promote protein-protein association, which is in accord with several experiments
\cite{BeheBJ1978,RalstonBiophysChem1997,WillsBiochemJ1983,Linse07PNAS}.

Although this qualitative argument is compelling, 
the molecular consequences of crowding effects on oligomer formation is unknown. In particular it is unclear whether crowding particles stabilize ordered $\beta$-sheet oligomers, which might facilitate
amyloid formation, or stabilize amorphous aggregates with little $\beta$-sheet content, which would likely impede or delay amyloid formation. To address
these unresolved issues we have carried out molecular simulations of systems composed of either
two, three, or four TTR peptides to assess the effect of size of spherical NPs on the structural 
and energetic properties of TTR oligomers. We find that spherical  NPs destabilize
ordered dimer aggregates but stabilize trimer and tetramer ordered aggregates.  Increasing the 
radius of the NP, at a fixed crowder volume fraction ($\Phi_C$), reduces the effect of molecular
crowding due to an increase in the interstitial space between NPs. In contrast, 
increasing the aspect ratio of the NPs with a spherocylinder shape
leads to greater destabilization of the ordered dimer than a spherical crowding particle.

In order to assess the effect of crowders on oligomer formation we first carried out separate constant temperature simulations on the monomer, dimer (denoted $\{TTR\}_2$), trimer
($\{TTR\}_3$), and tetramer ($\{TTR\}_4$) systems in the EEF1.1 implicit solvent model \cite{Karplus1999} at $\Phi_C=0$, a temperature of 395 K, and at a protein concentration
around 30 mM (SI Table 1). A high temperature was used to increase conformational sampling and yield converged results; we found
that simulations carried out at temperatures less than 330 K were not converged. It should be stressed that for
a complete understanding of crowding effects peptide concentration must also be varied 
\cite{thirumPNAS2005} as has been recently demonstrated in a complementary study using 
model systems \cite{Pellarin10JPhysChemLett}.  For the systems containing two or more peptides we find that due 
to the high protein concentration the peptides are associated. Therefore, we cannot 
examine the effect of NPs on the association process of these peptides. However, 
as mentioned previously, the more crucial aspect from the perspective of amyloid formation 
is determining the effect of NPs on the different aggregated species (ordered versus disordered), 
an issue which these simulations can directly address.

The TTR monomer is compact and devoid of persistent
secondary structure at $\Phi_C=0$. Its average radius of gyration is 7 \AA, whereas a fully extended TTR
structure would have a value of around 12 \AA. Analysis of the secondary structure content shows that coil
and turn structures are the most prevalent while helical and $\beta$-strand structures are negligibly
populated ($<$ 2\%). Even in the presence of crowding agents the strand content does not change appreciably. Thus, in isolation the TTR peptide is unstructured.

\begin{figure}[t]
  \begin{center}
    \includegraphics[width=0.8\textwidth]{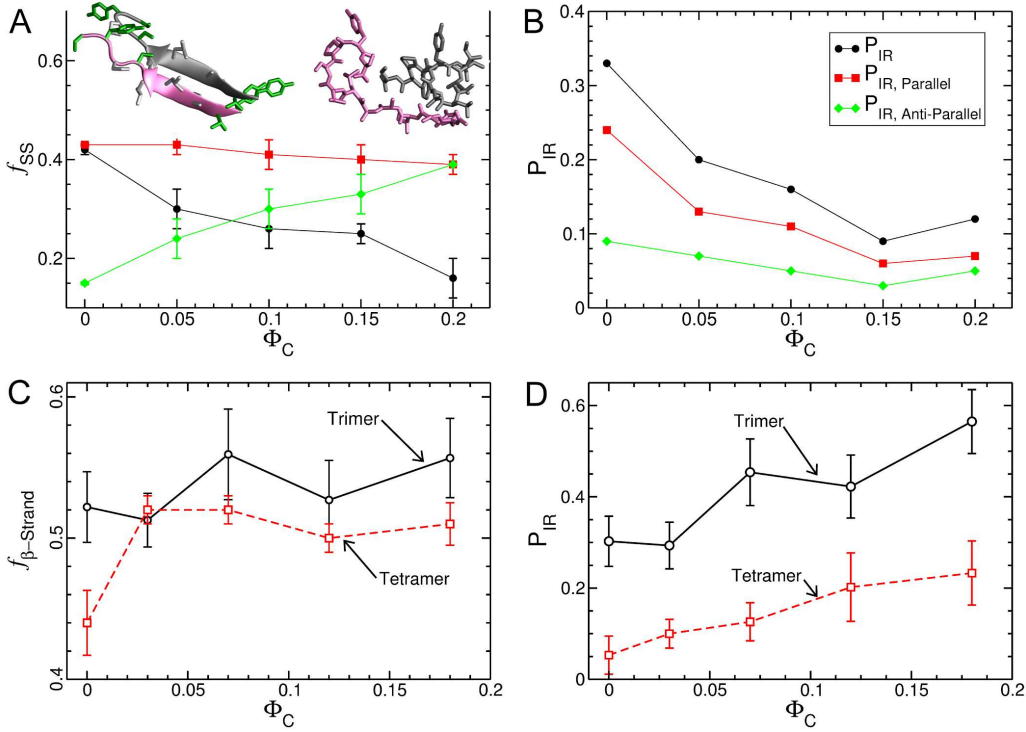}
  \caption{Effect of spherical crowding on $\{TTR\}_n$ as a function $\Phi_C$.
(A) and (B) are for the $\{TTR\}_2$. (A) Secondary structure content ($\beta$-strand (black), random coil (red), or
turn(green)) versus $\Phi_C$. Example structures from the ordered aggregate and disordered aggregate
are displayed on the upper left and right respectively. (B) The fraction of in-register contacts ($P_{IR}$, see Eq. 4 in
the SI) versus $\Phi_C$. (C) and (D) correspond to data from the $\{TTR\}_3$ and $\{TTR\}_4$ systems.
(C) The $\Phi_C$-dependent changes in the $\beta$-strand content for the trimer (solid black line with circles)
and the tetramer (dashed red line with squares).(D) Same as (C) except the ordinate is the probability
of finding an in-register aggregate (including both parallel and anti-parallel configurations).}
\label{dtt}
   \end{center}
\end{figure}

The dimer $\{TTR\}_2$ has an appreciable population of 
both ordered and disordered aggregates. In the ordered
state the peptides prefer parallel in-register $\beta$-sheets over anti-parallel structures.
The fraction of $\beta$-strand content in $\{TTR\}_2$ is 0.42 (Fig. \ref{dtt}a). 
The value of $P_{IR}$, the probability that the
two strands are in-register, when both
are in a parallel arrangement  is nearly 3-fold greater than the probability of forming
in-register anti-parallel structures (Fig. \ref{dtt}b). Expanded peptide conformations have several more in-register contacts on
average, make more inter-peptide backbone hydrogen bonds, and are more 
likely to have $\beta$-strand content than compact peptide conformations. 
For example, the radius of gyration $\la R_g\ra$ of the individual
peptides in the ordered $\{TTR\}_2$
(`ordered' being defined as having greater than eight in-register contacts out of a possible eleven)
is 10.8 \AA\ while $\la R_g\ra$ of the peptide in the disordered $\{TTR\}_2$ conformations (which have less than 5 
in-register contacts) is 6.7 \AA. 

Independent simulations of the $\{TTR\}_3$ and $\{TTR\}_4$  at $\Phi_C=0$ reveal 
they also exist as either disordered aggregates or
ordered $\beta$-sheets. However, the probabilities of $\beta$-sheet structure, which includes out-of-register $\beta$-strands, for $\{TTR\}_3$ and 
$\{TTR\}_4$ are 0.52 and 0.44, respectively (Fig. \ref{dtt}c). 
The probability that $\{TTR\}_3$ and $\{TTR\}_4$ form ordered 
in-register parallel and anti-parallel strand arrangements
are 0.30 and 0.06, respectively (Fig. \ref{dtt}d). 
\begin{wrapfigure}{r}{0.5\textwidth}
  \vspace{-20pt}
  \begin{center}
    \includegraphics[width=0.4\textwidth]{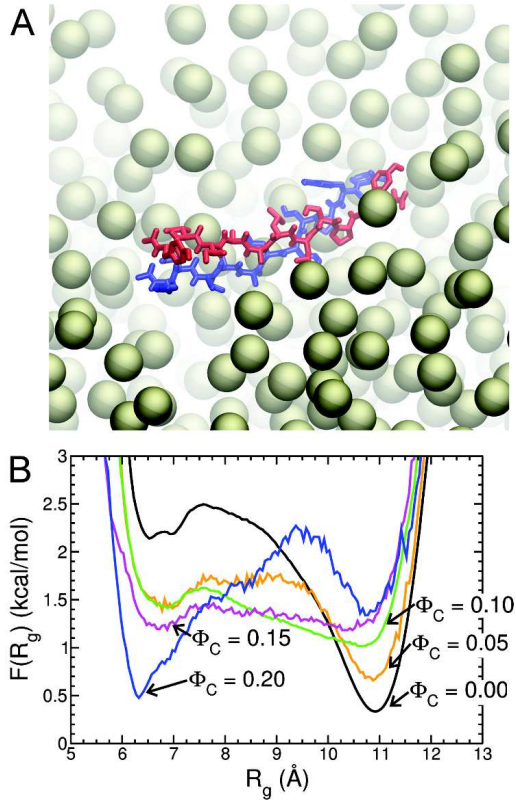}
  \end{center}
  \vspace{-20pt}
  \caption{Effect of spherical crowding on the size of the $\{TTR\}_2$ dimer.
(A) Snapshot of a single configuration of the NPs during the simulations. (B)
The free energy profile as a function of the radius-of-gyration
of the monomers in the dimer at various $\Phi_C$ values.}
  \label{np}
  \vspace{-10pt}
\end{wrapfigure}
The decrease in ordered in-register $\beta$-sheets, relative to the
$\{TTR\}_2$ system, is due to an increase in the number of 
low energy 
out-of-register disordered conformations that are accessible as the number of peptides in the aggregate increases. 
Thus, the landscape of TTR oligomers has a number of distinct basins of attraction with conformationally heterogeneous structures
(see also \cite{NguyenPNAS2007}). 

We next examined the influence of 
inert spherical NPs with $R_C=3.5$ \AA\  ($\Phi_C>0$, Fig. \ref{np}a). For the $\{TTR\}_2$ system
the presence of spherical NPs decreases the stability of the expanded (conformations with $R_g > 10$ \AA) 
dimer structures monotonically as $\Phi_C$ increases, while 
the stability of the collapsed structures (i.e., $R_g < 7.5$ \AA) increases (Fig. \ref{np}b). As a consequence, 
the average solvent accessible surface area and the molecular volume of the dimer decreases as 
$\Phi_C$ increases (data not shown). In addition, the 
$\beta$-strand content (Fig. \ref{dtt}a) and the probability of finding polymerization competent 
in-register dimers ($P_{IR}$) decreases with increasing $\Phi_C$ (Fig. \ref{dtt}b). 
Thus, relative to $\Phi_C=0$, crowding destabilizes in-register $\beta$-sheet aggregates causing
the equilibrium to shift towards collapsed associated dimer
structures at all values of $\Phi_C$  (see the structures in Fig. \ref{dtt}a).

Surprisingly, we find the opposite result for the $\{TTR\}_3$ and $\{TTR\}_4$, whose 
ordered structures  are stabilized as $\Phi_C$ increases. The $\beta$-strand content in $\{TTR\}_3$  remains the same as $\Phi_C$ increases,
whereas there is a slight increase for $\{TTR\}_4$ (Fig. \ref{dtt}c). In addition, $P_{IR}$
increases for the trimer and tetramer with increasing $\Phi_C$ (Fig. \ref{dtt}d).
The observed variations in crowding-induced changes in the stabilities of the ordered $\beta$-sheet aggregates may be
relevant to \emph{in vivo} amyloid formation as $\beta$-sheets are the primary structural component
found in mature fibrils \cite{PettyPNAS2005,EisenbergNature2005,TyckoME2006,TarusJACS2006}. 
In particular, NPs may inhibit the  ordered dimer structures 
but stabilize higher order oligomers. 

This non-monotonicity in the $n$-dependence  
of the stability of the ordered aggregate of $\{TTR\}_n$ ($n=2,3,4$) at various $\Phi_C$ can be explained using a theory based on
depletion forces.  Scaled Particle Theory \cite{CotterJCP1970,PierottiCR1976,MintonBBA2003} 
suggests that a factor influencing crowder effects is
the free energy cost of creating a void within the configuration of NPs to accommodate $\{TTR\}_n$ when 
$\Phi_C\neq 0$. The free energy is a function of volume of $\{TTR\}_n$'s volume, which increases as $n$ becomes larger. 
Clearly, the probability of finding a volume to accommodate $\{TTR\}_n$ decreases as $n$ increases.  
Depletion-induced changes in the free energy of the ordered aggregate 
relative to the disordered aggregate, denoted $\Delta G_n$, is therefore proportional to
$ln(V_{OA} - V_{DA})$, where $V_{OA}$ ($V_{DA}$) is the volume 
excluded to the NP's center-of-mass (also known as the `depletion volume')
by the ordered (disordered) aggregate. 
Peptide conformational entropy changes also likely contribute to the behavior of $\Delta G_n$ with $n$.
However, it is difficult to accurately estimate these entropy changes in the ordered and disordered aggregates
due to crowding.
In addition to conformational entropy differences, stability is also determined by favorable enthalpic inter-peptide
interactions. Enthalpic stabilization occurs for both the ordered and disordered aggregates as $n$ increases 
because of an increase in the number of inter-peptide contacts. The specific 
value of $n$ which stabilizes the ordered form should depend on the size and sequence of the peptide.
For $\{TTR\}_n$ it is only when $n$ becomes larger than 3 that the ordered aggregate structures become 
more stable in the presence of spherical NPs (Figs. \ref{dtt}b and \ref{dtt}d). 

If depletion forces are a major factor in  the non-monotonic behavior of $\Delta \Delta G_n$ 
with $n$ then we can account for the simulation results using the changes in the 
volume excluded to the NPs.  From the Asakura-Oosawa (AO) theory \cite{OosawaJCP1954} and a microscopic formulation \cite{ShawPRA1991}, as
well as several other approaches \cite{CotterJCP1970,PierottiCR1976,MintonBBA2003},  it follows that the difference
in free energy between species $i$ and $j$ is
 $\Delta G_i(n) = P(V_{ex}^i(n) - V_{ex}^{j}(n))$, 
where $P=k_BT\Phi_C / V_C$, is the osmotic pressure under ideal solution conditions, $k_B$ is Boltzmann's 
constant, $V_C$ is the molecular volume of a nanoparticle, and $V_{ex}^i(n)$ is the 
volume excluded to the spherical nanoparticle by species $i$. There  are three relevant species in 
the $\{TTR\}_n$ system; the soluble non-associated monomers (SM), the disordered aggregate (DA), and the ordered aggregate (OA).
We calculate $V_{ex}^i(n)$ by assuming that 
the soluble monomers and disordered aggregate are hard  spheres whose radii 
depend on $n$. In this case 
\begin{eqnarray}
V_{ex}^{SM}(n) &=& \frac{4\pi n}{3}[(1.927N_{aa}^{0.6}+R_C)^3-(1.927N_{aa}^{0.6})^3] \label{soluble} \\
V_{ex}^{DA}(n) &=& \frac{4\pi}{3}[(1.927(n\cdot N_{aa})^{0.6}+R_C)^3-(1.927(n\cdot N_{aa})^{0.6})^3] \label{disordered},
\end{eqnarray}
where $N_{aa}$ is the number of amino acids in the peptide and the exponent of 0.6 is the Flory scaling
exponent that characterizes  the size of  a protein in a good solvent \cite{FloryPPCBook}.
The shape of the ordered $\beta$-sheet aggregate is taken to be a stacked rectangular parallelepiped,
where each parallelepiped corresponds to one TTR peptide. For this species $V_{ex}$ is
computed using
\begin{eqnarray}
V_{ex}^{OA}(n) &=& (l+2R_C)(w+2R_C)(h+2R_C)-lwh-(8R_C^3-\frac{4}{3}\pi R_c^3), \label{beta_sheet}
\end{eqnarray}
where $l$, $w$, and $h$ correspond to the length, width and
height of a $\beta$-sheet made up of $n$ peptides, respectively. Therefore, we define 
$l = 3.2\cdot N_{aa}$ \AA, corresponding to the length of an extend 
$\beta$-strand made up of $N_{aa}$ amino-acids; $w = 8$ \AA\,  which corresponds to the
distance that the side chains stick out above and below the $\beta$-sheet;  $h = 4.85 n$  \AA, where $4.85$ \AA $ $ corresponds to the experimentally determined distance between the  neighboring strands in a $\beta$-sheet.

\begin{wrapfigure}{r}{0.5\textwidth}
  \vspace{-20pt}
  \begin{center}
    \includegraphics[width=0.4\textwidth]{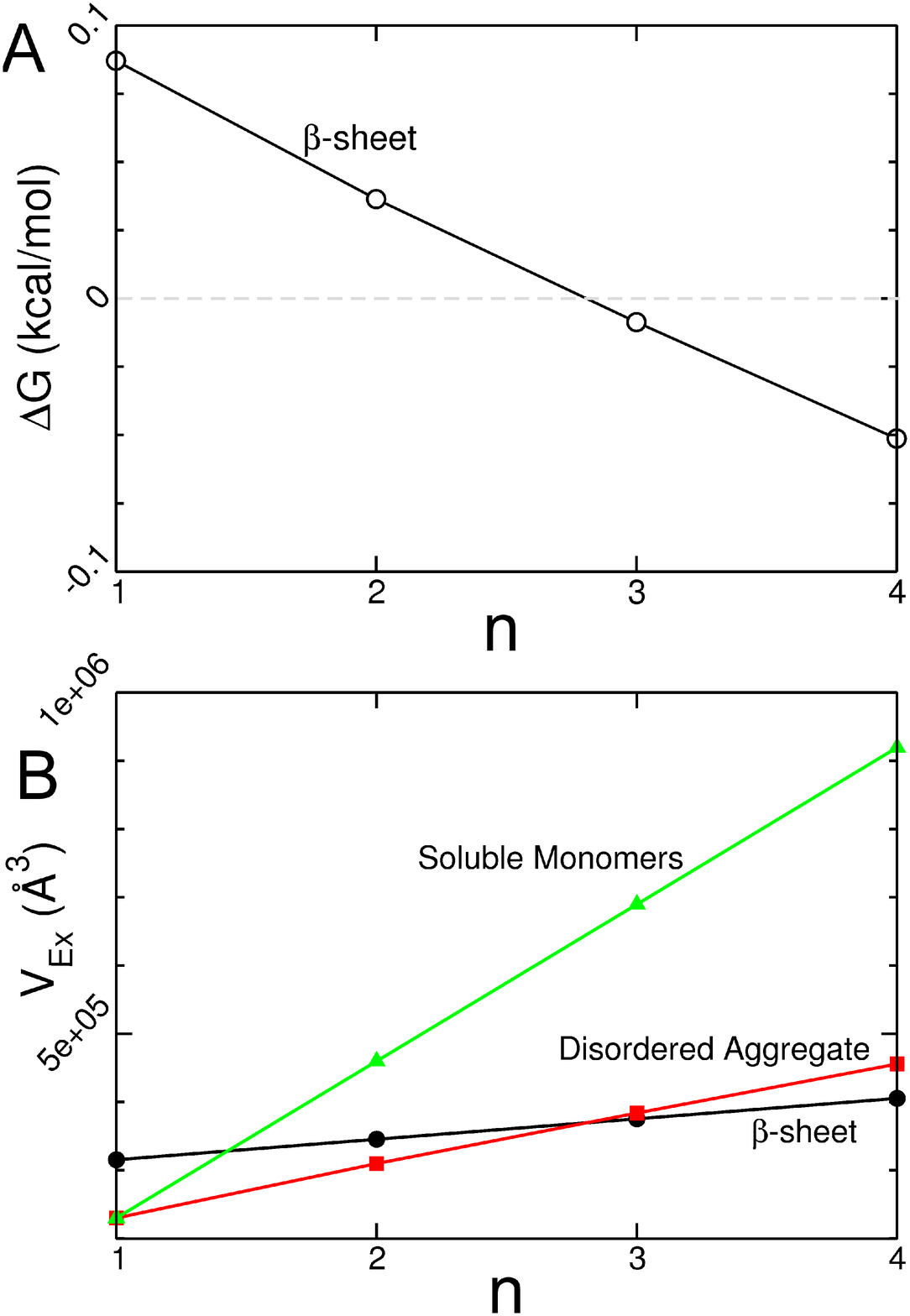}
  \end{center}
  \vspace{-20pt}
  \caption{(A) Theoretical prediction of the stability of the ordered $\beta$-sheet aggregate relative to the disordered aggregate as a
function of the number of TTR peptides in the system. The solution conditions in this
theoretical treatment are T = 395 K, $\Phi_C$=0.15, and $R_C$=30 \AA.
(B) Volume excluded (i.e., the depletion volume) to the NPs center of mass of the
soluble monomers (green triangles), disordered aggregate (red squares) and ordered $\beta$-sheet
aggregate (black circles) calculated from Eqs. \ref{soluble}-\ref{beta_sheet}}
  \label{ao}
  \vspace{-10pt}
\end{wrapfigure}

Using Eqs. \ref{soluble}-\ref{beta_sheet} it is straightforward to compute the $V_{ex}$  for each species
at different $\Phi_C$, and thereby estimate $\Delta G_i(n)$ at a given $\Phi_C$. We find that indeed
for a range of $R_C$ values this model displays non-monotonic behavior (Fig. \ref{ao}a) consistent with the trends observed in the simulations; the ordered dimer
is destabilized by spherical crowders, but as more peptides are added the NPs stabilize the ordered
$\beta$-sheet aggregates. The physical origin of this unusual behavior arises because disordered dimer aggregate excludes less volume
than the ordered dimeric $\beta$-sheet (Fig. \ref{ao}b), but as additional peptides are added to the system the volume per peptide
increases to a greater extent in  the disordered aggregate than the ordered $\beta$-sheet (the
same observation was made in ref. \cite{RohrigBJ2006}). 
As a consequence,
the  free energy of the system  is minimized when the dimer is disordered
whereas NP-induced ordered trimer and tetramer structures are more stable.  Recapitulation of the
qualitative trends observed in the simulation model suggests that the depletion
effect explains the changes in nanoparticle-induced changes in stability as $n$ varies.

\begin{wrapfigure}{r}{0.5\textwidth}
  \vspace{-20pt}
  \begin{center}
    \includegraphics[width=0.4\textwidth]{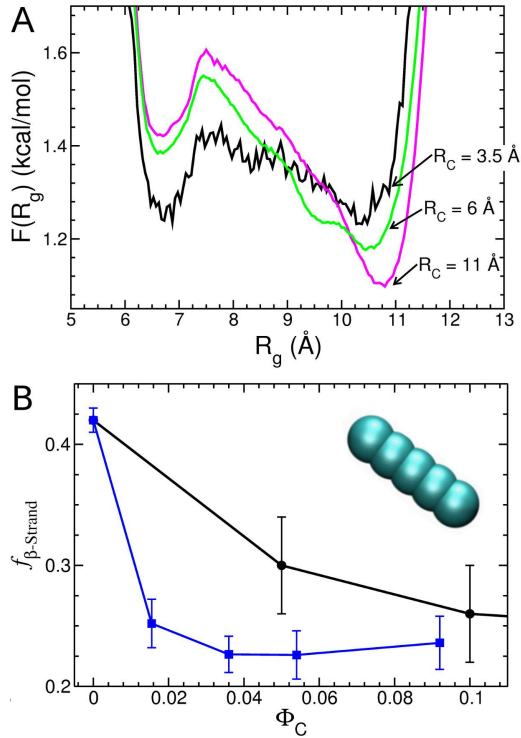}
  \end{center}
  \vspace{-20pt}
  \caption{(A) Effect of spherical NP size on $\{TTR\}_2$ at $\Phi_C$ = 0.15.
Changes in the free energy profile as a function of $R_g$ of the monomer (within the dimer) as the radius $R_C$ of the spherical NP is varied.
(B) Average $\beta$-sheet content of the $\{TTR\}_2$ system versus $\Phi_C$ for spherical
crowders (black line and circles), and spherocylindrical crowders with an aspect ratio of 3.3 (blue line and squares). The
inset shows a spherocylinder used in the simulations.}
  \label{rc}
  \vspace{-10pt}
\end{wrapfigure}

There are a number of other parameters besides $\Phi_C$ that also influence oligomerization. What happens when the radius of the NPs increases? For a fixed peptide concentration ([$C_P$]) and $\Phi_C$,
larger NPs exclude less volume to the protein than smaller crowders because the interstitial space between crowders increases
with crowder size. It can be shown that the strength of the depletion force is proportional to $\frac{1}{R_C^2}$. 
Therefore, larger crowding particles should have
a smaller effect in altering the stability of $\{TTR\}_n$. Explicit simulations that we ran of $\{TTR\}_2$ at $\Phi_C = 0.15$ using
three spherical NP
sizes ($R_C = 3.5,$ 6 and 11 \AA) show that indeed as $R_C$ increases
larger crowders shift the equilibrium  towards expanded structures just as
 observed in bulk simulations (Fig. \ref{rc}). 
However, in the simulations we observe that
the average $\beta$-sheet content does not exhibit an increase 
as the size of the NP increases and the probability ($\approx 0.06$) 
of finding ordered in-register dimers  is also unchanged. Thus, increasing
$R_C$ of spherical NPs decreases the impact of crowding - a result consistent with other theoretical and simulation
modeling \cite{BatraBJ2009}.

Another variable that controls the extent or ordered oligomer formation is the shape of the crowding particles.  
Indeed, \emph{in vivo}  most naturally occurring molecular crowders are unlikely to be spherical. 
Analysis of protein structures in the Protein Data Bank \cite{DimaJPCB2004} show that even in the folded state proteins adopt 
anisotropic shapes. To explore the effect of NP shape on dimer stability we simulated the $\{TTR\}_2$
system in the presence of spherocylindrical NPs in the isotropic phase whose aspect ratio (=$L/D+1$, where $L$ is the spherocylinder
length and $D$ is its diameter) is 3.3. We find that for the same $\Phi_C$ value (below $\Phi_C$=0.1, which is below the isotropic to nematic transition point) 
spherocylindrical NPs cause a greater loss of $\beta$-strand content in the dimer than spherical NPs. Thus, spherocylindrical 
NPs destabilize ordered oligomers. This
suggests that anisotropic NPs may have a greater impact on amyloid formation than spherical NPs, and that
crowder shape is another important variable in understanding amyloid formation \emph{in vivo}. To further explore realistic models of oligomerization in the cellular context
it would be important to consider a soup of proteins as was done recently to explore folding and diffusion in models of \textit{E. Coli.} \cite{Elcock10PLOS}.

Although the systems investigated in this study are caricatures of cellular crowding they illustrate the complexity of 
peptide aggregation under \emph{in vivo} conditions. While generic arguments suggest that depletion forces should promote protein association,
our work shows that the structures of the oligomers can be dramatically altered by the size, shape, and 
volume fraction of the  nanoparticles. Our study shows that an interplay of a number of factors determines the 
equilibrium between the ordered and disordered oligomeric structures. 
For spherical NPs the variables that determine the depletion forces, 
and hence the strength and range of the entropically-induced inter peptide attraction, are 
$q = \frac{R_g}{R_C}$ (where $R_g$ is the size of the peptide monomer), [$C_P$] the peptide concentration, 
and $\Phi_C$. 

This predicted complexity in the crowding-induced diagram of states of 
the amyloidogenic peptide is not surprising given that in the field of polymer physics
it was discovered previously that 
the phase diagram of a mixture of hard sphere colloidal particles 
and self-avoiding polymers in an athermal solvent is complicated,
and depends on polymer concentration, $\Phi_C$ and $q$ 
\cite{Fleer08AdvCollIntSci}. For the TTR peptides and $R_C$ values 
examined here $q$ ranges in value from 0.6 up to 2 (Note that $R_g \approx$7 \AA\ for the peptide monomer). 
It follows from our work that larger $q$ 
values can promote extended peptide conformations with high $\beta$-strand 
content in aggregates consisting of more than two peptides. Conversely,
in the extreme case of $q << 1$ (in the so called colloid limit 
\cite{Fleer08AdvCollIntSci}) we predict that the influence of crowding effects will be decreased,
and the equilibrium between ordered and disordered species may not be significantly perturbed relative to 
bulk ($\Phi_C$=0).

The results presented here show that understanding of NP effects on protein aggregation 
requires determination of a global phase diagram in terms of a number of variables.  
The non-monotonicity in NP-induced changes in stability should be contrasted with the effect 
of crowding on monomeric protein folding stability, which monotonically  increases in the presence of 
non-adsorbing crowding particles \cite{thirumPNAS2005}. Further 
insights into  \emph{in vivo} aggregation will require 
extension of the present work by taking crowder phase and 
polydispersity into account perhaps along the lines suggested 
recently \cite{ElcockCOSB2010}. In such a complex
environment polydispersity and concentration fluctuations could 
drive phase separation in the macromolecules \cite{Walter95FEBS}, 
which could add additional complications in the process of protein aggregation.
\\
\\
\noindent
\textbf{\large{Methods}}
\\
\\ 
\textit{Models for peptide and solvent.}
We chose the  peptide fragment (sequence Tyr-Thr-Ile-Ala-Ala-Leu-Leu-Ser-Pro-Tyr-Ser) from the Transthyretin protein (TTR) 
corresponding to residues 105-115. The structure of these 12 residues in an amyloid fibril has been determined (PDB ID 1RVS) using 
solid state NMR \cite{GriffinPNAS2002,GriffinPNAS2004}.
We cap the peptide's N-terminus with an acetyl blocking group and its C-terminus with
an amine blocking group ($NH_2$). For the peptide we use an all-atom representation except for non-polar hydrogen atoms, which are omitted
in the calculations.
The EEF1.1 implicit solvent model \cite{Karplus1983} is used to account for solvent effects in conjunction
with the CHARMM19 force-field \cite{Karplus1999}.

\textit{The EEF1.1 implicit solvent model.}
The parameterization of the EEF1.1 \cite{Karplus1999} force-field requires the use of specific cutoff distances for non-bonded interactions. The Lennard-Jones (LJ) and electrostatic interactions are
truncated at 9 \AA, with a switch function applied to the LJ term starting at 7 \AA. The Lorentz-Berthelot mixing rules \cite{TildesleyBook} are used to determine the undefined $\sigma$-values
between the atomic centers in the peptide. A distance dependent
dielectric constant is used for electrostatic interactions.

The LJ interactions in the EFF1.1 protein parameters are restricted to a distance less than 9 \AA, 
which prevents us from studying crowders with $R_C$ values greater than 4.5 \AA\ in the standard Charmm
code. We modified the CHARMM code to allow the LJ cutoff
to be dependent on the identity of the interacting atomic centers, thereby allowing us to use larger crowding particles. For  protein-protein interactions
nonbonded interactions are truncated at 9 \AA, but for crowder-crowder and crowder-protein interactions we use a cutoff of $2R_C + 2$ \AA\ and $R_C + 2$ \AA, respectively. These procedures allow us to 
maintain the EEF1.1 nonbond requirements for protein-protein interactions while allowing for larger crowding particles to be simulated.

\textit{Simulation details.} 
Simulations were carried out in the NVT ensemble at 395
K using Langevin dynamics with a friction coefficient of 1 $ps^{-1}$.
This high temperature allowed the
the dimer simulations to reach equilibrium, whereas simulations
at 330 K were found to not have converged on our simulated time scales (data not shown).
The SHAKE algorithm was used to fix the bond lengths of covalently bonded hydrogen,
allowing the use of a 2 fs time-step.

Typically, we generated ten independent trajectories for each crowder size and $\Phi_C$. At least half of the trajectories were started with peptide conformations taken
from equilibrated bulk simulations at $\Phi_C = 0.0$. When possible the other 
five starting conformations were taken from solution conditions that were 
closest to those of interest. For example, when simulating crowders with 
$R_C = 6$ and $11$ \AA, half of the initial protein conformations were from 
the equilibrium simulations of crowders with $R_C = 3.5$ \AA\ and 
$\Phi_C$ = 0.15. 

A box length of 60.0 \AA\ to 80.0 \AA\ is used for all spherical NP
conditions, which gives a peptide concentration in the range from 15 mM to 31 mM. At $\Phi_C = 0.05$, 0.10, 0.15, 0.20
there are 60, 120, 180 and 240 crowders in the periodic box, respectively. Because
correlations in the pair distribution function between crowders do not persist
for more than half the box length, finite size effects are minimized (Fig. S1).
A summary of the simulation details can be found in Table S1.

\textit{Models for spherical particles.}
Spherical NPs are modeled as Lennard-Jones particles.
The interaction between  sites $i$ and $j$ on two distinct NPs is
\begin{eqnarray}
V_{LJ}(r_{ij}) &=& 4\epsilon\left[  \left(  \frac{\sigma} {r_{ij}}\right)^{12} -
\left(\frac{\sigma}{r_{ij}} \right)^6 \right]. \label{vnbond_attr}
\end{eqnarray}
The choice of $\epsilon = 10^{-15}$ kcal/mol makes Eq. \ref{vnbond_attr} essentially repulsive,
and the condition $V_{LJ}(r_{ij}=2R_C) = k_BT$, allows us to solve for $\sigma$.
We consider three values for $R_C = 3.5,6,$ and $11$ \AA, which leads to $\sigma$= 108.7, 186.3, and 341.6 \AA, respectively.

\textit{Model for spherocylinders and crowder-protein interactions.}
We define all protein-crowder $\sigma$-values such that
when a crowding particle interaction site is ($R_C$ + 1) \AA\ from a protein atomic center $V_{LJ}(R_C+1) = k_BT$.
The resulting $\sigma$ values are 139.7, 217.4 and 372.6 \AA\ for $R_C$ = 3.5, 6 and 11 \AA, respectively.

The covalent bond between the tethered spheres that comprise the spherocylinder was modeled using
\begin{eqnarray}
V_{B}(r) &=& K_B/2 (r_o - r)^2 \label{vbond}
\end{eqnarray}
where $K_B=30$ kcal/(mol$\cdot$\AA$^2$) and $r_o=4$ \AA\ is the equilibrium bond length.
The rigidity of the spherocylinder is maintained using a bond angle potential
\begin{eqnarray}
V_{A}(\theta) &=& \frac{K_A}{2} (\theta_o - \theta)^2\label{vangle}
\end{eqnarray}
with $\theta_{o} = 180^o$ and $K_A = 0.6092$ $kcal/degree^2$. An important characteristic of a spherocylinder is
the aspect ratio $\lambda$. A value of $\lambda=1$ corresponds to a spherical crowder with diameter $D=2R_C$. In our study
all spherocylinders have $R_C=3.5$ \AA, and $L_C$, which is
proportional to the number of spheres ($N_B$) in the spherocylinder. $L_C = (N_B-1)4.0$ \AA. We used
$\lambda = 3.3,$ corresponding to a spherocylinder made up of 5 beads.
\\
\\
\noindent
\textbf{Acknowledgments}
This work was supported by a NIH grant (R01 GM076688-080) to JES and DT, a NSF postdoctoral Fellowship to EO, and the intramural program
at the National Heart Lung and Blood Institute of the NIH to BB.
\\
\\
\noindent
\textbf{Supporting Information}
Analysis details. This material is available free of charge via the Internet at http://pubs.acs.org.

\newpage
\noindent
\textbf{\large{References}}
\providecommand*{\mcitethebibliography}{\thebibliography}
\csname @ifundefined\endcsname{endmcitethebibliography}
{\let\endmcitethebibliography\endthebibliography}{}

\end{document}